# Investigation of Scattering Property for An Anisotropic Dielectric Circular Cylinder

Li Ying-Le[1], Li Jin[2]

1 Box399, Institute of Radio Wave Propagation & Scattering, Xianyang Normal University, 712000, China

2 School of Science, Xidian University, 710071, China

E-mail: liyinglexidian@yahoo.com.cn

Utilizing the scales theory of electromagnetic theory, the anisotropic dielectric material is reconstructed into an isotropic medium. The analytic expressions of scattering field and the scattering breadth for an anisotropic material cylinder are first presented. Their validities are checked theoretically. The influences induced by the dielectric constant tensor etc. on the scattering breadth are simulated. The results show that the scatterings both in the forward direction and vertical direction to the incident direction are strong. The dielectric constant in the polarizing direction has a biggish effect on scattering field. The mechanism of results is presented.

**PACS**：41.20.Jb，42.25.Dd，42.25.Fx

## Ⅰ. INTRODUCTION

The investigations of interactions between electromagnetic (E.M.) waves and the anisotropic dielectric targets and their applications have become increasingly important in recent years, particularly in the areas of remote-sensing and detection [1-6]. The E.M. wave's propagation in the uniaxial anisotropic medium is researched [7] by using the transmission matrix, however this wave propagation in a general anisotropic dielectric material could not be studied. In literatures [8][9], The E.M. wave scattering features of a two dimensional anisotropic material and expanded form of wave function in the material are analyzed in detail. The numerical algorithms, for instance the moment method, are also utilized to study the E.M. scattering of three dimensional anisotropic medium targets [10]. From the means of studying, the techniques used in researching the scattering of E.M wave by the anisotropic targets may be divided into two kinds, namely, the analytical method and the numerical method. The analytical research is not very appeared since its complexity. The results of E.M. propagation and scattering in part literatures are incorrect [15] since the dielectric constant tensor having been considered as the same both in the right angle coordinate system and spherical coordinate system and neglect the factor that the orthogonality of wave functions is based on the wave equation that obtained in the isotropic space.

In this paper, we use the scales theory of E.M. wave to reconstruct the anisotropic material into an isotropic medium. The analytical expressions of E.M. waves inside and outside the anisotropic cylinder are presented and then its scattering width is obtained. The influences of dielectric tensor and the incident wave angle etc. on scattering width are demonstrated. The obtained results are useful in the anisotropic target's detection and the time harmonic factor $e^{-j\omega t}$ is adopted.

## II. THE SCATTERING PROPERTY OF AN ANISOTROPIC DIELECTRIC CIRCULAR CYLINDER

A Electromagnetic reforming of an anisotropic dielectric cylinder

There is an anisotropic medium cylinder, which radius is $R$. Its center and the original of the primary coordinate system $\Sigma$ are located at the same point. Its axis is in z-direction as shown in Fig.1 and electromagnetic parameters are given as

$$\boldsymbol{\varepsilon} = \varepsilon_r \varepsilon_0 = \varepsilon_0 \begin{bmatrix} \varepsilon_1 & 0 & 0 \\ 0 & \varepsilon_2 & 0 \\ 0 & 0 & \varepsilon_3 \end{bmatrix} \quad \mu = \mu_0 \tag{1}$$

We know from literatures [11][12] that the electromagnetic parameters in $\Sigma$' coordinate system become

$$\boldsymbol{\varepsilon}' = \begin{bmatrix} \dfrac{\varepsilon_1 bc}{a} & 0 & 0 \\ 0 & \dfrac{\varepsilon_2 ac}{b} & 0 \\ 0 & 0 & \dfrac{\varepsilon_3 ba}{c} \end{bmatrix} \quad \boldsymbol{\mu}' = \begin{bmatrix} \dfrac{\mu_0 bc}{a} & 0 & 0 \\ 0 & \dfrac{\mu_0 ac}{b} & 0 \\ 0 & 0 & \dfrac{\mu_0 ba}{c} \end{bmatrix}$$

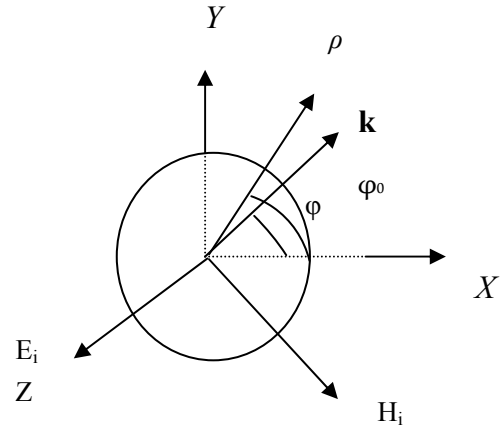

**Figure1: A plane electromagnetic wave irradiates a cylinder**

The symbols $a, b$ and $c$ are the scales' factors. In order to reduce the error, we may choose the factors as

$$a = \frac{m^2}{\sqrt{\varepsilon_2 \varepsilon_3}}, \quad b = \frac{m^2}{\sqrt{\varepsilon_1 \varepsilon_3}}, \quad c = \frac{m^2}{\sqrt{\varepsilon_2 \varepsilon_1}}, \quad m = \sqrt{\frac{\varepsilon_1 + \varepsilon_2 + \varepsilon_3}{3}}$$

$$\boldsymbol{\varepsilon}' = \begin{bmatrix} m^2 & 0 & 0 \\ 0 & m^2 & 0 \\ 0 & 0 & m^2 \end{bmatrix} \quad \boldsymbol{\mu}' = \begin{bmatrix} \dfrac{\mu_0 m^2}{\varepsilon_1} & 0 & 0 \\ 0 & \dfrac{\mu_0 m^2}{\varepsilon_2} & 0 \\ 0 & 0 & \dfrac{\mu_0 m^2}{\varepsilon_3} \end{bmatrix} \quad (1)$$

Expression (1) follows that the electromagnetic anisotropic material may be reformed into a isotropic dielectric medium and anisotropic material in magnetism in Σ' coordinate system. Since $\mu_0 = 10^{-9}$ and $m^2/\varepsilon_i$ has the same order, so the difference in magnetism becomes the order of $10^{-10}$. This error may be neglect in applications, so we can think that the cylinder in Σ' coordinate system is an isotropic material one.

B The scattering property of an anisotropic dielectric cylinder

A plane electromagnetic wave polarizes in z-direction. Its expression in Σ coordinate system is

$$\mathbf{E}_{iz} = E_0 \exp(-jk_0 \rho \cos(\varphi - \varphi_0)) = E_0 \sum_{n=-\infty}^{n=\infty} j^{-n} J_n(k_0 \rho) e^{-jn(\varphi - \varphi_0)} \quad (2)$$

From the Maxwell's equation $\nabla \times \mathbf{E} = -j\omega\mu_0 \mathbf{H}$, we have

$$H_\phi = \frac{1}{j\omega\mu_0} \frac{\partial E_z}{\partial \rho}$$

It is so obtained

$$H_{\varphi i} = -\frac{jE_0}{\eta_0} \sum_{n=-\infty}^{n=\infty} j^{-n} J_n'(k_0 \rho) e^{-jn(\varphi - \varphi_0)} \quad (3)$$

According to the incident wave expressions, the scattering fields are assumed as

$$E_{sz} = E_0 \sum_{n=-\infty}^{n=\infty} j^{-n} a_n H_n^{(2)}(k_0 \rho) e^{-jn(\varphi - \varphi_0)} \quad (4)$$

$$H_{\varphi s} = -\frac{jE_0}{\eta_0} \sum_{n=-\infty}^{n=\infty} j^{-n} a_n H_n^{(2)'}(k_0 \rho) e^{-jn(\varphi - \varphi_0)} \quad (5)$$

The inner field of the cylinder in Σ' coordinate system may be written as

$$E'_{tz} = TE_0 e^{-jmk_0 \rho' \cos(\varphi' - \varphi_0')} \quad (6)$$

After expressing the angle function in (6) with the functions in Σ coordinate system[11][12], we have

$$\cos(\varphi' - \varphi_0') = \frac{\sqrt{A^2 + B^2}}{a^2 b^2 g g_0} \cos(\varphi - \beta_0)$$

After putting above expression into expression (6), we obtain the expression in Σ coordinate

system

$$\mathbf{E}_{tz} = T\frac{E_0}{c}e^{-jk_t\rho\cos(\varphi-\beta_0)} = \frac{E_0}{c}\sum_{n=-\infty}^{n=\infty}j^{-n}b_nJ_n(k_t\rho)e^{-jn(\varphi-\beta_0)}$$

$$A = b^2\cos\varphi_0 \quad B = a^2\sin\varphi_0 \quad k_t = mk_0\frac{\sqrt{A^2+B^2}}{a^2b^2g_0} \tag{6}$$

$$\cos\beta_0 = \frac{A}{\sqrt{A^2+B^2}} \quad \sin\beta_0 = \frac{B}{\sqrt{A^2+B^2}} \quad g_0 = \left(\frac{\cos^2\phi_0}{a^2}+\frac{\sin^2\phi_0}{b^2}\right)^{\frac{1}{2}}$$

When the material is an isotropic one, that is $a = b = c = 1$, $m = \sqrt{\varepsilon}$, and now $\beta_0 = \phi_0, g_0 = 1, k_t = mk_0$, expression (6) has been changed as

$$\mathbf{E}_{tz} = E_0\sum_{n=-\infty}^{n=\infty}j^{-n}b_nJ_n(mk_0\rho)e^{-jn(\varphi-\phi_0)}$$

The above is just the expression used in the literatures for solving the scattering field from an isotropic dielectric cylinder. The validity is demonstrated here. By using the Maxwell equation and above formula, we may obtain the magnetic field as

$$H_{t\varphi} = \frac{-jE_0}{c\eta_0}\frac{k_t}{k_0}\sum_{n=-\infty}^{n=\infty}j^{-n}b_nJ_n'(k_t\rho)e^{-jn(\varphi-\beta_0)} \tag{7}$$

On the cylinder's surface $\rho = R$, the tangential components of **E** and **H** are continued. So we get

$$E_0\sum_{n=-\infty}^{n=\infty}j^{-n}J_n(k_0R)e^{-jn(\varphi-\varphi_0)} + E_0\sum_{n=-\infty}^{n=\infty}j^{-n}a_nH_n^{(2)}(k_0R)e^{-jn(\varphi-\varphi_0)} = \frac{E_0}{c}\sum_{n=-\infty}^{n=\infty}j^{-n}b_nJ_n(k_tR)e^{-jn(\varphi-\beta_0)}$$

$$-\frac{jE_0}{\eta_0}\sum_{n=-\infty}^{n=\infty}j^{-n}J_n'(k_0R)e^{-jn(\varphi-\varphi_0)} - \frac{jE_0}{\eta_0}\sum_{n=-\infty}^{n=\infty}j^{-n}a_nH_n^{(2)'}(k_0R)e^{-jn(\varphi-\varphi_0)} = \frac{-jE_0}{c\eta_0}\frac{k_t}{k_0}\sum_{n=-\infty}^{n=\infty}j^{-n}b_nJ_n'(k_t\rho)e^{-jn(\varphi-\beta_0)}$$

We simplify the above expressions and obtain

$$J_n(k_0R)e^{jn\phi_0} + a_nH_n^{(2)}(k_0R)e^{jn\phi_0} = b_n\frac{J_n(k_tR)}{c}e^{jn\beta_0} \tag{8}$$

$$J_n'(k_0R)e^{jn\varphi_0} + a_nH_n^{(2)'}(k_0R)e^{jn\varphi_0} = \frac{k_t}{k_o}b_n\frac{J_n'(k_tR)}{c}e^{jn\beta_0} \tag{9}$$

From expressions (8)(9), the coefficients are easy derived

$$a_n = \frac{k_tJ_n'(k_tR)J_n(k_0R) - k_0J_n(k_tR)J_n'(k_0R)}{k_0J_n(k_tR)H_n^{(2)'}(k_0R) - k_tJ_n'(k_tR)H_n^{(2)}(k_0R)}$$

$$b_n = ck_0e^{-jn(\beta_0-\phi_0)}\frac{H_n^{(2)'}(k_0R)J_n(k_0R) - H_n^{(2)}(k_0R)J_n'(k_0R)}{k_0J_n(k_tR)H_n^{(2)'}(k_0R) - k_tJ_n'(k_tR)H_n^{(2)}(k_0R)} \tag{10}$$

Now the scattering fields and the transmitting fields of the anisotropic dielectric cylinder are obtained since the coefficients $a_n$ and $b_n$ are presented. It is obviously that when the target is isotropic medium $k_t = mk_0$, expression (10) is in agreement with that in the literature. We put the expression $H_n^{(2)}(x)$ into expression (4) as $x \gg 1$, the following is obtained

$$E_{sz} = E_0 \sqrt{\frac{2}{\pi k_0 \rho}} e^{-j\left(k_0\rho - \frac{\pi}{4}\right)} \sum_{n=-\infty}^{n=\infty} a_n e^{-jn(\varphi-\varphi_0)}$$

The scattering width is presented

$$k(\phi) = \frac{2}{\pi k_0} \left| \sum_{n=-\infty}^{n=\infty} a_n e^{-jn(\varphi-\varphi_0)} \right|^2 \qquad (11)$$

When the incident wave polarizes in x-direction, we may also research the scattering property of anisotropic material cylinder. There is no need to study it here. Followings are partial simulations:

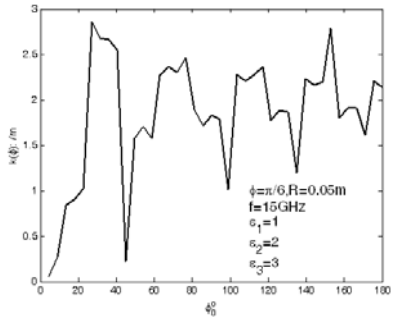

Figure 2：Scattering width changes with the incident angle

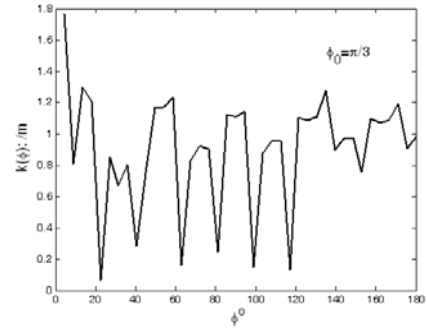

Figure 3：Scattering width changes with the observing angle

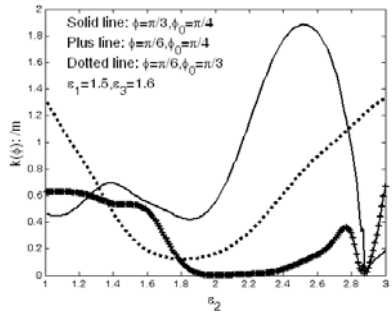

Figure 4: Scattering width changes with $\varepsilon_2$

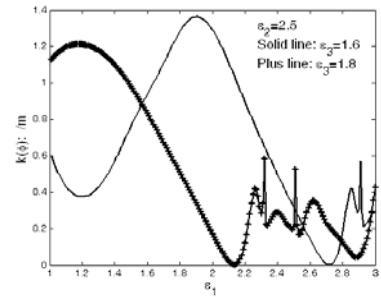

Figure 5：Scattering width changes with $\varepsilon_1$

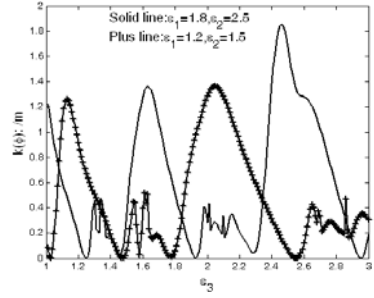

**Figure 6**: Scattering width changes with $\varepsilon_3$

In figure 2, the change of scattering width versus the incident angle is shown in which the observing angle is $\phi = \pi/6$. They are shown that for the anisotropic dielectric cylinder there is a buildup effect in the forward scattering direction, in the direction vertical to the incident direction, namely $\phi_0 = 2\pi/3$ or $\phi_0 = \pi/3$, the scattering width is bigger and $K(\varphi)$ generally shows a periodicity. Figure 3 shows the change of $K(\varphi)$ with the observing angle. Its periodicity is obvious. In the direction perpendicular to the incident direction, that is $\phi = 5\pi/3$, the scattering is big, which is similar to dipole radiation. Figure 4 to figure 6 demonstrate the influences induced by the dielectric tensor. It can be seen that the change of scattering width versus to the tensor is very complex. When the dielectric constant is in the direction vertical to the electric field, the change of scattering width versus to it is slowness. When the dielectric constant is in the direction parallel to the electric field, the change of scattering width versus to it is acuteness and there is a periodicity. Since we always measure the scattering field in the direction perpendicular to $z$-direction, this measured field should be the radiated field. It is mainly produced by electric dipole in $z$-direction. When the magnitude of incident wave is given, the electric dipole is proportional to the dielectric constant in $z$-direction. Therefore the dielectric constant in $z$-direction has a bigger influence.

## III. CONCLUSION

Based on the theory of electromagnetic field, we reconstruct the anisotropic dielectric material into isotropic one. The influence induced by dielectric tensor on permeability is analyzed after the reconstruction. By using the relations of parameters between the scale coordinate system and the primary coordinate system, the scattering field and the scattering width for an anisotropic medium cylinder are presented. The impacts resulted by the observing azimuth, wave incident angle and the tensors on the scattering width are researched. In Ku wave band, partial simulations are calculated. It is shown that in the forward scattering direction and in the direction vertical to the incident direction the scattering width is bigger. The dielectric constant in the polarizing direction has a large impact on scattering width and its physical analysis is processed. How to with the scales theory of electromagnetic field research the scattering properties of a finite anisotropic dielectric cylinder and the anisotropic medium sphere are our future works, which may fill up the shortness of no analytical solutions in E.M. wave scattering for the simple anisotropic material targets.

## IV. ACKNOWLEDGEMENTS

This work was supported by the National Natural Science Foundation of China (Grant Nos 60971079), the Natural Science Foundation of Shaanxi Province (Grant Nos 2009JM8020) and Natural Science Foundation of Shaanxi Educational Office (Grant Nos 09JK800)